\documentclass[twocolumn]{article}
\usepackage[T1]{fontenc}          
\usepackage{lmodern}              
\usepackage{textcomp}             
\usepackage{amsmath}
\usepackage{amssymb}
\usepackage{tabularx}
\usepackage{graphicx}
\usepackage{booktabs} 
\usepackage{cuted}
\usepackage{caption}
\usepackage{subcaption}
\usepackage{placeins}
\usepackage{url}      
\usepackage{hyperref}
\usepackage{titling}
\usepackage[backend=biber]{biblatex}
\usepackage{float}  
\usepackage{placeins}

\usepackage{tikz}
\usetikzlibrary{shapes, arrows, positioning, calc}

\begin{document}

\title{Adaptive Orchestration for Large-Scale Inference on Heterogeneous Accelerator Systems: Balancing Cost, Performance, and Resilience}
\author{%
  Yahav Biran, Imry Kissos  \\
  Amazon \\
  \vspace{0.8em}
  \textcolor{red}{github.com/aws-samples/scalable-hw-agnostic-inference}
}
\date{}
\maketitle

\section*{Abstract}
The surge in generative AI workloads has created a need for scalable inference systems that can flexibly harness both GPUs and specialized accelerators while containing operational costs. This paper proposes a hardware-agnostic control loop that adaptively allocates requests across heterogeneous accelerators based on real-time cost and capacity signals. The approach sustains low latency and high throughput by dynamically shifting between cost-optimized and capacity-optimized modes—ensuring the most efficient use of expensive compute resources under fluctuating availability. Evaluated using the Stable Diffusion model \cite{stabilityaisd}, the framework consistently meets latency targets, automatically redirects traffic during capacity shortfalls, and capitalizes on lower-cost accelerators when possible. These results highlight how a feedback-driven deployment strategy, spanning the entire software and hardware stack, can help organizations efficiently scale generative AI workloads while maintaining resilience in the face of limited accelerator capacity.

\section{Introduction}
The growing demand for generative applications in AI has driven advancements in compute accelerators, leading to a range of solutions designed to enhance performance and manage costs. Accelerators such as NVIDIA GPUs and purpose built chips like AWS Inferentia and Trainium now play a central role in enabling efficient inference at scale. However, integrating these diverse hardware architectures into existing workflows remains a challenge, requiring careful consideration of inference modes such as eager-based and graph-based, software compatibility, workload optimization, and cost-performance trade-offs.

Inference, the process of using trained machine learning models to generate predictions or outputs, is distinct from training in its emphasis on dynamic, latency-sensitive operations. For generative AI applications like text-to-image models, this requirement becomes even more critical. These models demand a balance of high computational throughput and low response times, as shown in \ref{fig:breakpoint-latency}, particularly in scenarios where user interactions are time-sensitive. At the same time, cloud-based inference must remain cost-efficient and adaptable to fluctuating availability of compute resources.
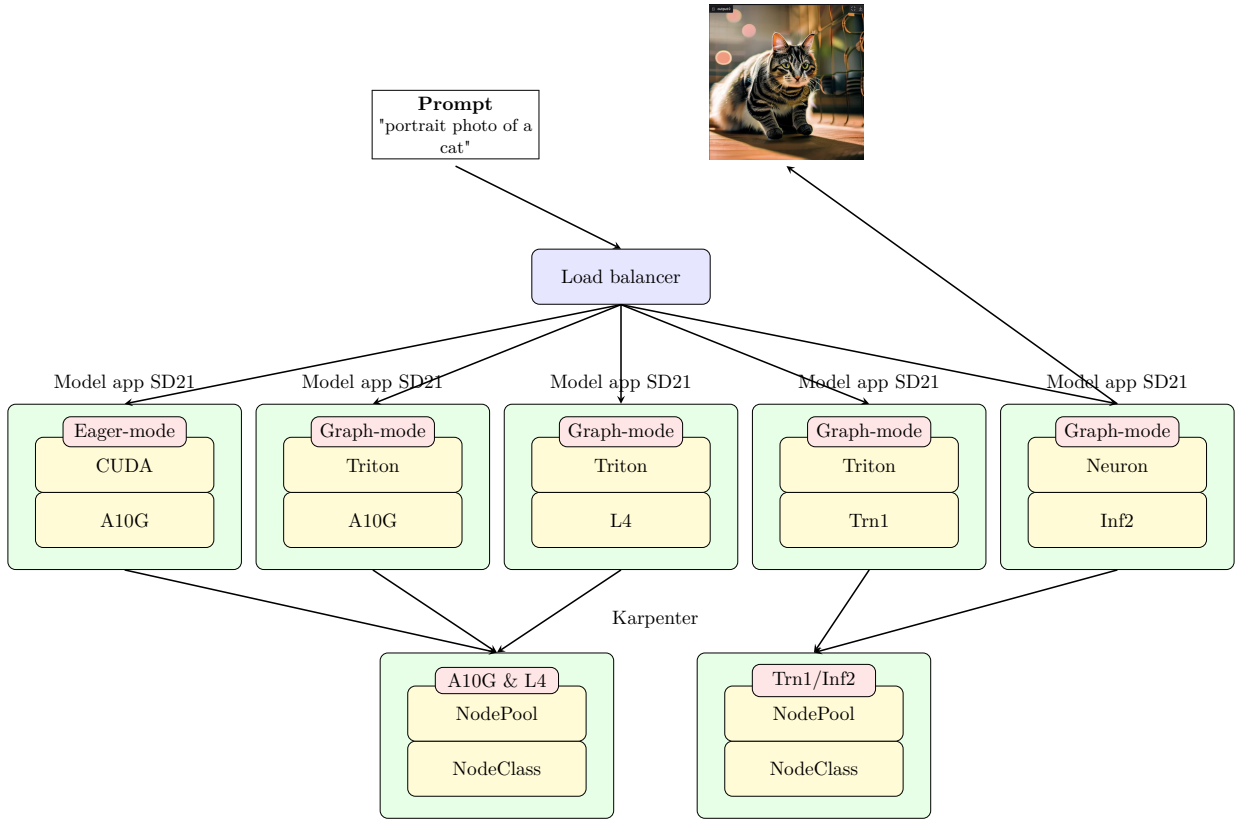
\begin{figure*}[ht]
    \centering
    \resizebox{1.0\textwidth}{!}{  
        \begin{tikzpicture}[
            node distance=1.8cm and 5cm,
            mainbox/.style={
                rectangle,
                draw,
                rounded corners,
                align=center,
                text width=3cm,
                fill=blue!10,
                minimum height=1cm
            },
            subbox/.style={
                rectangle,
                draw,
                rounded corners,
                align=center,
                text width=4cm,
                fill=green!10,
                minimum height=3cm
            },
            subsubbox/.style={
                rectangle,
                draw,
                rounded corners,
                align=center,
                text width=3cm,
                fill=yellow!20,
                minimum height=1cm
            },
            subsubsubbox/.style={
                rectangle,
                draw,
                rounded corners,
                align=center,
                text width=2cm,
                fill=red!10,
                minimum height=0.275cm 
            },
            arrow/.style={
                thick,
                ->,
                >=stealth
            }
        ]
        \node[mainbox] (LB) {Load balancer};
        \node[anchor=south] (InPrompt) at ([yshift=1.5cm,xshift=-3.0cm]LB.north) {\fbox{\parbox{2.8cm}{\centering
          \textbf{Prompt}\\\small "portrait photo of a cat"
        }}};
        \node[anchor=south] (OutImage) at ([yshift=1.5cm,xshift=3.0cm]LB.north) {\includegraphics[width=2.8cm]{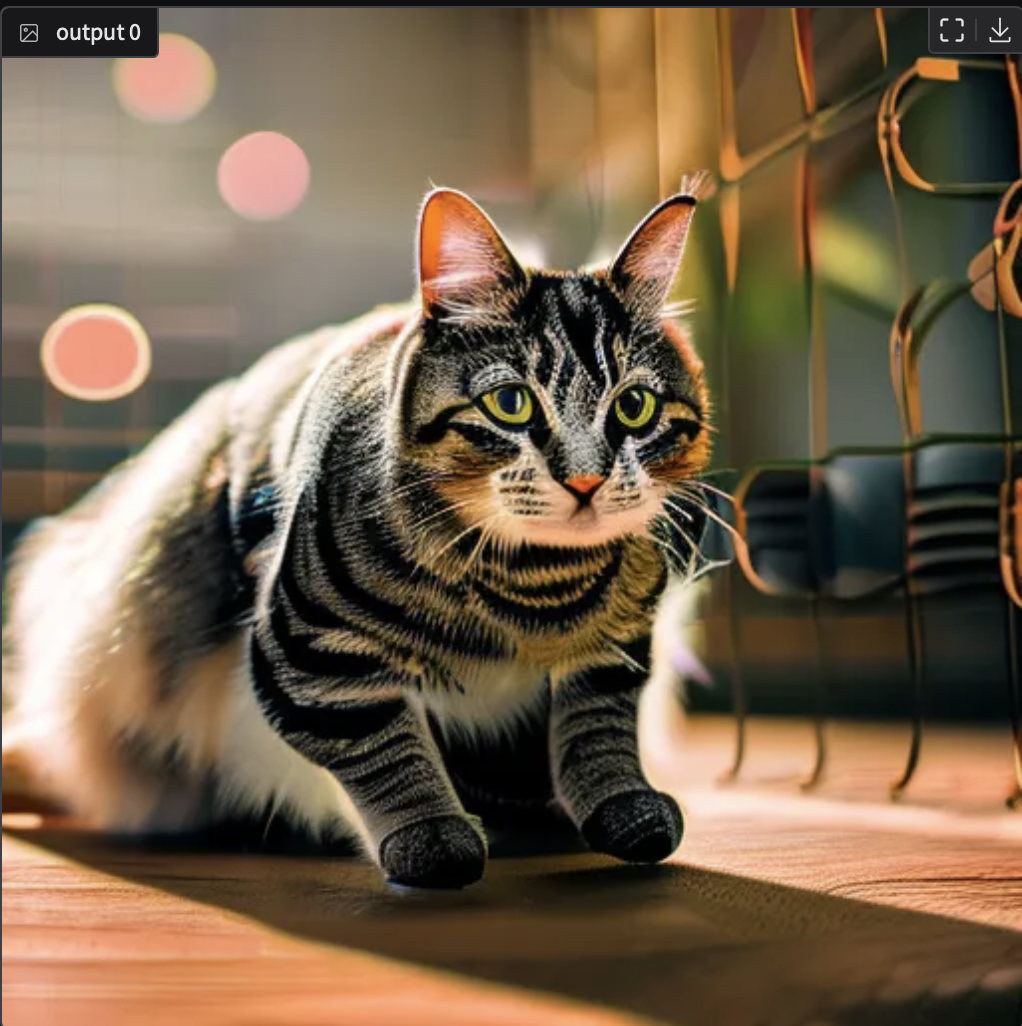}};
        \draw[arrow] (InPrompt.south) -- (LB.north);

        \foreach \i/\xpos/\hw/\fw/\mode in {
            App1/-9cm/A10G/CUDA/Eager-mode,
            App2/-4.5cm/A10G/Triton/Graph-mode,
            App3/0cm/L4/Triton/Graph-mode,
            App4/4.5cm/Trn1/Triton/Graph-mode,
            App5/9cm/Inf2/Neuron/Graph-mode
        } {
            \node[subbox, below=of LB, xshift=\xpos, anchor=north] (\i) {};
            \node at (\i.north) [above=0.1cm] {Model app SD21};
            \node[subsubbox] at ([yshift=-0.6cm]\i.center) {\hw};
            \node[subsubbox] at ([yshift=0.4cm]\i.center) {\fw};
            \node[subsubsubbox] at ([yshift=1cm]\i.center) {\mode};
            \draw[arrow] (LB.south) -- (\i.north);
        }

        \draw[arrow] (App5.north) -- (OutImage.south);

        \node[subbox, below=1.5cm of App3, anchor=north, xshift=-2.25cm] (NewApp4) {};
        \node[subbox, right=1.5cm of NewApp4] (NewApp5) {};

        \node at ($(NewApp4.north east)!0.5!(NewApp5.north west)+(0,0.6)$) {Karpenter};
        \node at (NewApp4.north) [above=0.1cm] {};
        \node at (NewApp5.north) [above=0.1cm] {};

        \foreach \i/\pool/\class in {NewApp4/{A10G \& L4}/NodeClass, NewApp5/{Trn1/Inf2}/NodeClass}{
            \node[subsubbox] at ([yshift=-0.6cm]\i.center) {\class};
            \node[subsubbox] at ([yshift=0.4cm]\i.center) {NodePool};
            \node[subsubsubbox] at ([yshift=1cm]\i.center) {\pool};
        }

        \draw[arrow] (App1.south) -- (NewApp4.north);
        \draw[arrow] (App2.south) -- (NewApp4.north);
        \draw[arrow] (App3.south) -- (NewApp4.north);
        \draw[arrow] (App4.south) -- (NewApp5.north);
        \draw[arrow] (App5.south) -- (NewApp5.north);
        \end{tikzpicture}
    }
    \caption{The diagram illustrates a load balancer distributing traffic to five parallel model applications labeled "Model app SD21". Each application consists of compute framework and hardware accelerator components. The first two apps use CUDA (eager-mode) and Triton (graph-mode) with A10G accelerators. The third uses Triton with L4. The fourth and fifth run on Triton and Neuron frameworks, powered by Trn1 and Inf2 accelerators, respectively. Karpenter provisions resources through NodePools and NodeClasses managing NVIDIA (A10G, L4) and Neuron (Trn1, Inf2) instances. Arrows show the hierarchical relationships.}
\end{figure*}

This work presents a framework for scalable and hardware-agnostic inference that leverages Kubernetes-based\cite{kubernetes} tools to dynamically allocate resources across heterogeneous compute architectures. By integrating AWS Elastic Kubernetes Service (EKS), Kubernetes Event-Driven Autoscaling (KEDA)\cite{autoscaler_tool}, and Karpenter\cite{node_provisioner_tool}, this approach enables a flexible and efficient deployment of generative AI models across different accelerators, such as GPUs, AWS Inferentia, and Trainium. While this implementation highlights Kubernetes and specific open-source tools, the methods described are not limited to these technologies. The principles demonstrated can be extended to other container orchestration platforms and equivalent model invocations methods. Similarly, the framework is compatible with alternative compute hardware, including devices from vendors like AMD and Intel, beyond the specific accelerators evaluated here.

The framework adopts two key deployment strategies:
\begin{itemize}
    \item \textbf{Cost-Optimized Configuration:} This strategy minimizes costs by prioritizing accelerators with lower inference costs, adjusting traffic distribution accordingly.
    \item \textbf{Capacity-Optimized Configuration:} This approach ensures resilience by automatically switching to alternative accelerators during capacity constraints while maintaining latency and throughput requirements.
\end{itemize}
These strategies are evaluated in the context of Stable Diffusion, a generative AI model for text-to-image synthesis. Performance and cost metrics are analyzed across multiple compute accelerators that are invoked in various ways such as PyTorch eager-mode with GPU, graph-mode with GPU and XLA, demonstrating the effectiveness of the framework in managing heterogeneous workloads.

The primary contributions of this work are as follows:

\begin{itemize}
    \item A scalable framework for integrating and managing heterogeneous compute accelerators in inference workflows, adaptable to various container orchestration platforms and compute architectures.
    \item A detailed evaluation of Stable Diffusion performance on various accelerators, including NVIDIA GPUs and AWS AI chips, Inferentia and Trainium.
    \item Practical guidance for deploying and optimizing generative AI workloads in cloud environments using AWS services and Kubernetes tools.
\end{itemize}
This framework offers a practical pathway for organizations to deploy generative AI applications for serving inference requests effectively, balancing the competing demands of cost, performance, and resource availability in diverse computing environments.

\section{Background and Related Work}

\textbf{Transformer Architecture \& Foundation Models}
The emergence of the Transformer architecture has fundamentally reshaped both natural language processing and computer vision. Models such as GPT-3 and the Vision Transformer (ViT) have demonstrated that self-attention-based architectures can scale effectively across modalities, enabling state-of-the-art performance in text generation, image classification, and multimodal reasoning \cite{vaswani2017attention,brown2020language,dosovitskiy2021vit}. As these models continue to grow in size and adoption, scalable and latency-aware inference infrastructure has become critical for deploying them in production environments.

\textbf{Hugging Face Transformers}
To address this challenge, several software building blocks have emerged to simplify and abstract the deployment of Transformer-based models. Hugging Face’s \texttt{transformers} library provides the widely used \texttt{pipeline} API, which bundles pretrained models with tokenizers and pre/post-processing steps for common tasks. While this abstraction accelerates prototyping and experimentation, it is designed for single-node execution and lacks orchestration features such as autoscaling, load balancing, or multi-accelerator support \cite{wolf2020transformers}.

\textbf{PyTorch Execution \& Backend Support}
At the execution layer, PyTorch plays a foundational role by supporting both eager-mode and graph-mode inference. With backends like CUDA for NVIDIA GPUs and XLA for AWS Trainium and Inferentia, PyTorch enables developers to deploy the same model across a variety of accelerators without changing core logic \cite{paszke2019pytorch,aws_neuron_pytorch}. These hardware-agnostic abstractions are essential for decoupling model development from infrastructure-specific execution details.

\textbf{Ray Serve}
Ray Serve offers a general-purpose serving framework that abstracts inference deployments into autoscaling actors running on Ray clusters. It supports composable DAGs, multi-model workloads, and heterogeneous infrastructure \cite{moritz2018ray}. However, while Ray Serve provides flexibility at the application layer, it does not natively handle cost-performance trade-offs or real-time hardware-aware scheduling—capabilities needed to fully optimize inference across diverse accelerators.

\textbf{vLLM and PagedAttention}
For high-throughput serving of large language models specifically, \texttt{vLLM} introduces several inference-time optimizations tailored to the autoregressive decoding pattern. Its core innovation, \textit{PagedAttention}, allows the key–value cache to be paged efficiently in GPU memory, reducing fragmentation and maximizing memory reuse. These techniques enable 2--4× higher throughput at equivalent latency compared to conventional implementations \cite{kwon2023vllm}. \texttt{vLLM} also supports tokenizer parallelism and asynchronous scheduling for efficient batching. While originally designed for GPU inference, \texttt{vLLM} now includes support for AWS Neuron devices, enabling deployment on Inferentia and Trainium hardware \cite{vllm_neuron_support}. However, its focus remains on LLMs, and it does not yet extend to vision or multimodal models.

Our work builds on these capabilities by introducing an adaptive orchestration framework that coordinates deployment across heterogeneous accelerators based on real-time cost, latency, and capacity signals. Unlike Hugging Face pipelines or PyTorch backends, which offer single-device execution, and unlike systems like \texttt{vLLM} or Ray Serve, which focus on either intra-model efficiency or generic scaling, our approach fills the critical gap of cross-model, cross-hardware orchestration. By dynamically shifting between cost-optimized and capacity-optimized deployment strategies, our system achieves robust, efficient inference under variable load and constrained accelerator availability—delivering a unified, cloud-native control loop for managing generative workloads at scale.

\section{Optimization Framework}
\subsection{Problem Formulation} 
A deployment unit, denoted as \( DU_i \), is characterized by a triplet  coupled to  \((model, hardware, framework)\). This unit represents an AI application that has been deployed and is ready to be invoked via a REST interface (such as HTTP) or a remote call (like gRPC). The \( model \) component can be one of several machine learning models, including Llama, StableDiffusion, CLIP, or YOLO. The \( hardware \) component specifies the underlying infrastructure, which could be Inferentia2 instances (\( Inf2 \)), Trainium1 instances (\( Trn1 \)), or NVIDIA based instances such as $A10G$, $L4$, and $A100$. Each deployment unit is designed to operate within a single device, ensuring optimal performance and resource utilization. The \( framework \) refers to the software environment that facilitates the execution of the \( model \) on the specified \( hardware \). Examples of such frameworks include Neuron as PyTorch graph mode for \( Inf2 \) and \( Trn1 \), and PyTorch eager mode with CUDA or TorchDynamo graph-based mode like Triton for \( A10G \) and NVIDIA \( L4 \) chips.

The system's performance can be quantified using several key metrics. The cumulative number of inference requests to the model deployed in \( DU_i \) at a given time \( t \) is represented by \( N^{model}(t) \). The throughput of \( DU_i \) for a batch size of one, denoted as \( T_i(t) \), is calculated as the rate of change of \( N^{model}_i(t) \) with respect to time, mathematically expressed as:
\[
T_i(t) = \frac{dN^{model}_i(t)}{dt}
\]

Availability and demand for deployment units are managed through specific metrics. The maximum number of \( DU_i \) available in the pool at time \( t \) is represented by \( DU^p_i \), while the number of \( DU_i \) requested at the same time is denoted by \( DU^r_i \).

Throughput demand and supply at any given time \( t \) are critical for ensuring system efficiency. The throughput demand, \( T^d(t) \), corresponds to the total number of samples that need to be processed, denoted as \( N^{modelDemand}_i(t) \), and is defined by:
\[
T^d(t) = \frac{dN^{model}_i(t)}{dt}
\]
Conversely, the throughput supply, \( T^s(t) \), represents the total number of samples processed by the system, denoted as \( N^{modelProcessed}_i(t) \), and is similarly defined by:
\[
T^s(t) = \frac{dN^{model}_i(t)}{dt}
\]

Latency, denoted as \( L_i \), measures the end-to-end time taken to process a single sample on a deployment unit. This includes the time required to read input data from memory and write the processed output data back to memory, ensuring a comprehensive assessment of the system's responsiveness.

Finally, the deployment cost per time interval \( \Delta t \), represented as \( DU^c_i \), quantifies the financial expenditure associated with allocating a \( DU_i \) during the specified interval \( \Delta t \). This metric is essential for budgeting and optimizing resource allocation within the deployment environment.

\subsection{Optimization Goal}
We aim to request \( DU^r_i \) (\( i = 1, 2, 3, 4, 5 \)) to minimize the total cost, defined as:

\begin{equation}
\text{Minimize} \quad \sum_{i=1}^{5} DU^r_i(t) \cdot DU^c_i
\end{equation}

subject to the following constraints:

1. \textbf{Throughput Constraint:} The sum of throughput from all $DU^r_i$, $T^s(t)$, must meet or exceed the total throughput demand \( T^d(t) \):
\begin{equation}
T^s(t) = \sum_{i=1}^{5} DU^r_i(t) \cdot T_i \geq T^d(t)
\label{eq:throughputconstraint}
\end{equation}

2. \textbf{Capacity Constraint:} The requested $DU^r_i(t)$ must not exceed the available $DU^p_i(t)$ units:
\begin{equation}
DU^r_i(t) \leq DU^p_i(t), \quad \forall i \in \{1, 2, 3, 4, 5\}.
\label{eq:capacityconstraint}
\end{equation}

The average latency of the provisioned $DU^p$ units is given by:
\begin{equation}
L(t)\operatorname{avg} = \sum_{i=1}^{5} DU^r_i(t) \cdot L_i
\end{equation}

\subsection{Capacity Dynamics State Machine}
\label{subsec:capdynamics}
Ideally, user inference requests at the system ingress point can be efficiently handled by the $DU^r$ that satisfies the optimization goal (Equation 1), while other pools serve as a fallback when demand exceeds the cost-optimized resources (Equation 3). This approach balances efficiency and reliability. User inference requests are distributed with weights for each $DU^p_i$ that satisfy Equation 1. When the capacity constraint (Equation 3) is not satisfied, the system reduces the weight of $DU_i$ units lacking capacity and normalizes the distribution weight to meet the optimization goal using the remaining $DU^p$ units.

To simplify resource management, we categorize pools into two types:
\textbf{Cost-Optimized Weight:} Prioritizes cost efficiency, adhering to Equation 1; and \textbf{Capacity-Optimized Weight:} Distributes traffic across available capacity and requests new $DU^r_i$ based on availability.

This heuristic-based solution assumes cyclic workload distribution, balancing between cost and capacity-optimized configurations. The load resets at the end of a cycle, enabling transitions between two states without significant performance degradation.

The weight for each cost-optimized deployment unit is determined by the cost-to-latency ratio. Units with a lower cost relative to latency are assigned higher weights:

\begin{equation}\tag{5. Cost-Optimized Weights}
w^{cost}_i = \frac{\frac{1}{DU^c_i}}{\sum_{j=1}^{5} \frac{1}{DU^c_j}}
\end{equation}

Capacity-optimized weights are uniformly distributed among deployment units with available capacity:

\begin{equation}\tag{6. Capacity-Optimized (Uniform) Weights}
w^{cap}_i = \frac{1}{n}, \quad \text{for all } i \text{ where } DU^p_i(t) > 0
\end{equation}

The system switches between these two states based on Equation 2. The switching logic is:
\begin{strip}
\begin{equation}
w_i(t) =
\begin{cases} 
w_i^{\text{cost}}, & \text{if } \sum_{i=1}^5 DU^r_i(t) \cdot T_i \geq T^d(t) \text{ and } DU^r_i(t) \leq DU^p_i(t), \forall i, \\[10pt]
w_i^{\text{cap}}, & \text{if } \exists i : DU^r_i(t) > DU^p_i(t), \, n = |\{i : DU^p_i(t) > 0\}|.
\end{cases}
\end{equation}
\end{strip}

Adaptive throughput optimization is defined as:
\begin{strip}
\begin{equation}
T_s(t) =
\begin{cases} 
T_s^{\text{cost}}(t) = \sum_{i=1}^5 w_i^{\text{cost}} \cdot T_i \cdot DU^r_i(t), & \text{if cost-optimized weights are used}, \\[10pt]
T_s^{\text{cap}}(t) = \sum_{i=1}^5 w_i^{\text{cap}} \cdot T_i \cdot DU^r_i(t), & \text{if capacity-optimized weights are used}.
\end{cases}
\end{equation}
\end{strip}

\section{Methodology}
Our work focuses on enabling flexible, hardware-agnostic inference pipelines for machine learning models at scale. Building on the foundational principles described by recent research on distributed, compute-intensive systems, we adopt a containerized, cloud-native architecture that decouples model execution from underlying hardware resources. By thoughtfully combining containerization, automated orchestration, and performance-focused scaling strategies, we aim to deliver robust, cost-efficient inference across a range of heterogeneous computing environments\cite{k8shpc}.

\subsection{Designing a Hardware-Agnostic Architecture}
\textbf{Layered Abstraction}. We structured our system in layers to cleanly separate the complexities of data handling such as compute-accelerator's specific model graph, model execution, and resource orchestration. Inspired in part by approaches that standardize how models interact with the underlying hardware, we defined three conceptual tiers: (1) the \textbf{Data Plane}, which is responsible for handling incoming requests and preparing data inputs; (2) the \textbf{Model Execution Layer}, which encapsulates models within containerized runtime environments; and (3) the \textbf{Resource Orchestration Layer}, where scaling decisions and workload distributions are managed automatically.

This layered design ensures that any adjustments to hardware infrastructure—such as adding GPU-accelerated nodes or shifting to Neuron device-based nodes \cite{aws_ml_optimization} clusters—can occur without disrupting the model code itself as long as the performance metrics are met. Each containerized model runs inside a standardized environment, making it straightforward to port from one kind of hardware setup to another. This approach not only streamlines development and testing but also reduces operational overhead when deploying to different environments.

\subsection{Model Execution Layer}
Within this layer, we prioritize hardware-agnostic model execution by leveraging frameworks like PyTorch \cite{pytorch}. PyTorch, with its support for eager execution and graph modes, provides a flexible foundation for deploying models on diverse hardware accelerators.

\begin{itemize}
    \item Eager Mode: For rapid development and interactive experimentation, eager mode offers a familiar Python-like experience. This is particularly beneficial during the initial stages of model development and debugging.
    \item Graph Mode: When performance is critical, graph mode enables optimizations such as automatic differentiation and hardware-specific optimizations. This mode is crucial for achieving high throughput on accelerators like GPUs, AWS Inferentia, and Trainium.
\end{itemize}
PyTorch's abstraction layer plays a key role in dynamically discovering the available compute accelerator and selecting the most appropriate code path. For instance:

On NVIDIA GPUs, PyTorch can leverage TorchDynamo \cite{torchdynamo} for advanced optimizations, such as fusion and operator specialization.
On AWS Inferentia and Trainium, PyTorch can utilize pre-compiled kernels and optimized libraries for maximum performance.
This dynamic behavior allows developers to write code that seamlessly adapts to different hardware environments without significant code modifications. By abstracting away hardware-specific details, PyTorch empowers developers to focus on model development and optimization, while the framework handles the complexities of hardware-specific execution\cite{aws_trainium_systolic}.

Furthermore, within the Model Execution Layer, we employ containerization technologies like Docker and Kubernetes to package and deploy models as self-contained units. This approach facilitates portability and simplifies deployment across different environments. Each container includes all necessary dependencies, such as the model itself, the chosen PyTorch execution mode, and any required libraries or runtime environments.

By combining PyTorch's flexibility and hardware-agnostic capabilities with containerization technologies, we create a robust and portable foundation for deploying and managing machine learning models across a diverse range of hardware accelerators.\\
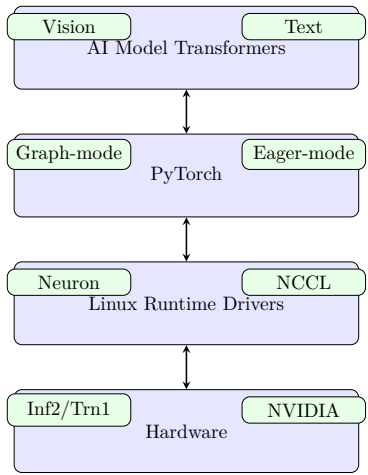
\begin{figure}[ht] 
\centering
\resizebox{0.3\textwidth}{!}{
\begin{tikzpicture}[
    node distance=0.8cm and 3cm, 
    mainbox/.style={
        rectangle,
        draw,
        rounded corners,
        align=center,
        text width=6cm,
        fill=blue!10,
        minimum height=1.5cm
    },
    subbox/.style={
        rectangle,
        draw,
        rounded corners,
        align=center,
        text width=2cm,           
        fill=green!10,
        minimum height=0.375cm     
    },
    arrow/.style={
        thick,
        <->,                      
        >=stealth
    }
]

    \node[mainbox] (AI) {AI Model Transformers};

    \node[subbox] (Vision) at ($(AI.north west)+(1cm,-0.375cm)$) {Vision};
    \node[subbox] (Text) at ($(AI.north east)+(-1cm,-0.375cm)$) {Text};

    \node[mainbox, below=of AI] (PyTorch) {PyTorch};

    \node[subbox] (GraphMode) at ($(PyTorch.north west)+(1cm,-0.375cm)$) {Graph-mode};
    \node[subbox] (EagerMode) at ($(PyTorch.north east)+(-1cm,-0.375cm)$) {Eager-mode};

    \node[mainbox, below=of PyTorch] (Linux) {Linux Runtime Drivers};

    \node[subbox] (GraphMode) at ($(Linux.north west)+(1cm,-0.375cm)$) {Neuron};
    \node[subbox] (EagerMode) at ($(Linux.north east)+(-1cm,-0.375cm)$) {NCCL};

    \node[mainbox, below=of Linux] (Hardware) {Hardware};
    \node[subbox] (GraphMode) at ($(Hardware.north west)+(1cm,-0.375cm)$) {Inf2/Trn1};
    \node[subbox] (EagerMode) at ($(Hardware.north east)+(-1cm,-0.375cm)$) {NVIDIA};

    \draw[arrow] (AI) -- (PyTorch);
    \draw[arrow] (PyTorch) -- (Linux);
    \draw[arrow] (Linux) -- (Hardware);

\end{tikzpicture}
}
    \caption{Model execution layer showing AI model, PyTorch, and supporting components.}
    \label{fig:model-exec-layer}  
\end{figure}

\subsection{Containerization and Model Packaging}
\textbf{Self-Contained Model Images}. Containerization plays a crucial role in deploying machine learning models by packaging each model with its dependencies, optimization libraries, and environment configurations into OCI-compliant containers \cite{oci_image_spec}. This approach ensures consistency across development and production, guaranteeing that models run as intended regardless of the underlying infrastructure. By leveraging OCI hooks, Docker can support accelerated compute environments such as NVIDIA GPUs and AWS AI chips. The oci-add-hooks tool injects prestart, poststart, and poststop hooks into a container’s config.json, exposing Inferentia and GPU devices to the containerized application. This method simplifies reproducibility and version control, enabling teams to deploy multiple model versions side by side or roll back to previous versions seamlessly. Each container serves as a fully reproducible snapshot, preserving the entire model environment at a specific point in time, which enhances scalability and streamlines the deployment of AI and ML workloads across heterogeneous hardware accelerators.

\subsection{Data Handling and Request Processing}
\textbf{Ingestion and Preprocessing}. At the Data Plane layer, we use RESTful endpoints to receive requests. Input payloads—whether they are images, or text—are normalized and preprocessed. This co-location of preprocessing code and model ensures that data transformations remain tightly coupled to the model’s logic.

We rely on AWS-native and HuggingFace\cite{huggingface_repos} services for data storage and retrieval of objects like model weights. The system seamlessly integrates object stores and HuggingFace model repositories, so changes to the underlying data infrastructure do not necessitate reworking the model deployment and invocation code.

\subsection{Automated Deployment and Validation}
\subsubsection{Build and Verification Pipelines} Before a model goes into production, we take it through an automated build and validation pipeline. This involves converting raw model artifacts—typically PyTorch or TensorFlow checkpoints—into production-ready inference engines such as TorchServe. As part of this step, we run basic performance tests and verify outputs against known inputs. If the model passes these checks, it’s packaged into a container and tagged as ready for deployment.

\subsubsection{Cluster-Oriented Deployment} For launching containers, we rely on Amazon EKS clusters. We choose the orchestration platform and hardware resources based on user-defined policies and current resource availability. Since we rely on standardized container interfaces, transitioning from Neuron-based nodes to GPU-based ones, or even combining the two, is straightforward with Karpenter’s Nodepool and Nodeclass abstractions. The system’s design aims to minimize the friction involved in leveraging specialized hardware accelerators.

\subsection{Dynamic Scaling and Resource Allocation}
\subsubsection{Elastic Orchestration and Scheduling} Once deployed, inference services run under a dynamic scaling regime. Using Kubernetes Horizontal Pod Autoscalers, the orchestrator continually monitors key performance indicators, such as request throughput, response latency, and resource utilization. When the load surges, it spins up additional container replicas that are powered by nodes that are launched by Karpenter\cite{node_provisioner_tool}. When demand subsides, it scales down gracefully. This responsive scaling strategy ensures we can meet performance targets without allocating unnecessary resources, ultimately helping reduce operational costs and fail-over to available capacity pools.

\subsubsection{Hardware-Agnostic Resource Policies} Unlike systems that hard-code certain models to specific hardware nodes, we rely on more generic requests and labels, allowing orchestrators to assign workloads to any suitable node. If GPUs are available and cost-effective, the scheduler takes advantage of them. If they’re not, it falls back to compatible and available instances. This kind of hardware abstraction aligns with our overall philosophy: the complexity of hardware selection is a background detail that the system handles, not a problem the developer or data scientist must solve each time.

\subsection{Observability and Continuous Improvement}
\textbf{Metrics, Logging, and Tracing}. To guide both automated decision-making and human-led optimization, we instrument the system with metrics that track latency, utilization, and throughput. We store logs and telemetry data using AWS CloudWatch, making it easy to correlate performance changes with code updates, configuration tweaks, or shifts in user behavior. This feedback loop helps us refine scaling policies, adopt new hardware types, and improve our preprocessing routines over time.
\textbf{Controlled Experiments and Benchmarking}. As we iterate on this approach, we conduct controlled experiments to measure improvements in efficiency, cost-effectiveness, and reliability. Drawing inspiration from the rigorous methodologies described in [1], we compare different scaling strategies, infrastructure configurations, and model optimization techniques. These experiments help us validate design decisions and confidently evolve our system as new hardware options and ML frameworks emerge.

\section{Experiments}
We conducted a series of experiments to evaluate the performance, scalability, and adaptability of our hardware-agnostic inference framework. Similar cost-latency trade-off studies \cite{heterogeneous_edge_eval,cost_efficient_gpu_cluster} confirm that. We designed our tests to answer the following key questions:

\begin{enumerate}
    \item \textbf{Performance Gains:} How does our containerized and hardware-agnostic approach compare to a baseline scenario using standard CPU-based instances (i.e., no hardware acceleration) in terms of latency, throughput, and cost efficiency?
    \item \textbf{Scaling Behavior:} How effectively does our autoscaling strategy respond to fluctuating workloads, and how does it compare to fixed-capacity or manually scaled deployments?
    \item \textbf{Hardware Adaptability:} Can the system seamlessly leverage a variety of accelerators, such as Amazon EC2 Inf2 and Trn1 instances running the AWS Neuron SDK, as well as NVIDIA GPUs (e.g., A10G, L4), without manual reconfiguration or performance regressions?
    \item \textbf{Quality Baseline:} Given the potential non-determinism in some ML model outputs, can the system maintain consistent output quality across different runs and hardware configurations?
\end{enumerate}

\subsection{Experimental Setup}

\textbf{Cluster Configuration.} We conducted our experiments on AWS environments capable of hosting a wide range of hardware options: Amazon EC2 Inf2 and Trn1 instances (accessed via the AWS Neuron SDK\cite{awsneuron}), and NVIDIA GPU-backed instances (A10G, L4)\cite{nvidia_hopper_arch}. For the baseline configuration, we used single deployment unit $DU^p_i$ of Amazon EC2 Inferentia2, Trainium1 and L4 and A10G NVIDIA GPU. We then used an heterogeneous mixtures of GPU, and Inferentia/Trainium instances. Cluster sizes ranged from small testbeds with a few instances to large-scale clusters with dozens of nodes and hundreds of container replicas.

\textbf{Models and Workloads.} We evaluated a variety of model architectures, including an image classification model, a Transformer-based text inference model, and a recommendation model. Additionally, to assess non-deterministic output quality, we used a generative model producing stylized images. Input request patterns ranged from stable, steady loads to highly variable, bursty traffic patterns resembling real-world demand fluctuations.

\textbf{Metrics and Instrumentation.} We measured end-to-end latency (95\textsuperscript{th} percentile), throughput (requests per second, RPS), resource utilization (CPU, GPU, Inferentia/Trainium), and estimated cost per 1,000 inferences. Logs and performance metrics were aggregated in Amazon CloudWatch, and traces were analyzed offline to understand container scheduling and autoscaling decisions.

\subsection{Quality Baseline Under Non-Determinism}
\begin{figure}[ht]
    \centering
    \begin{subfigure}[t]{\linewidth}
        \centering
        \includegraphics[width=\columnwidth]{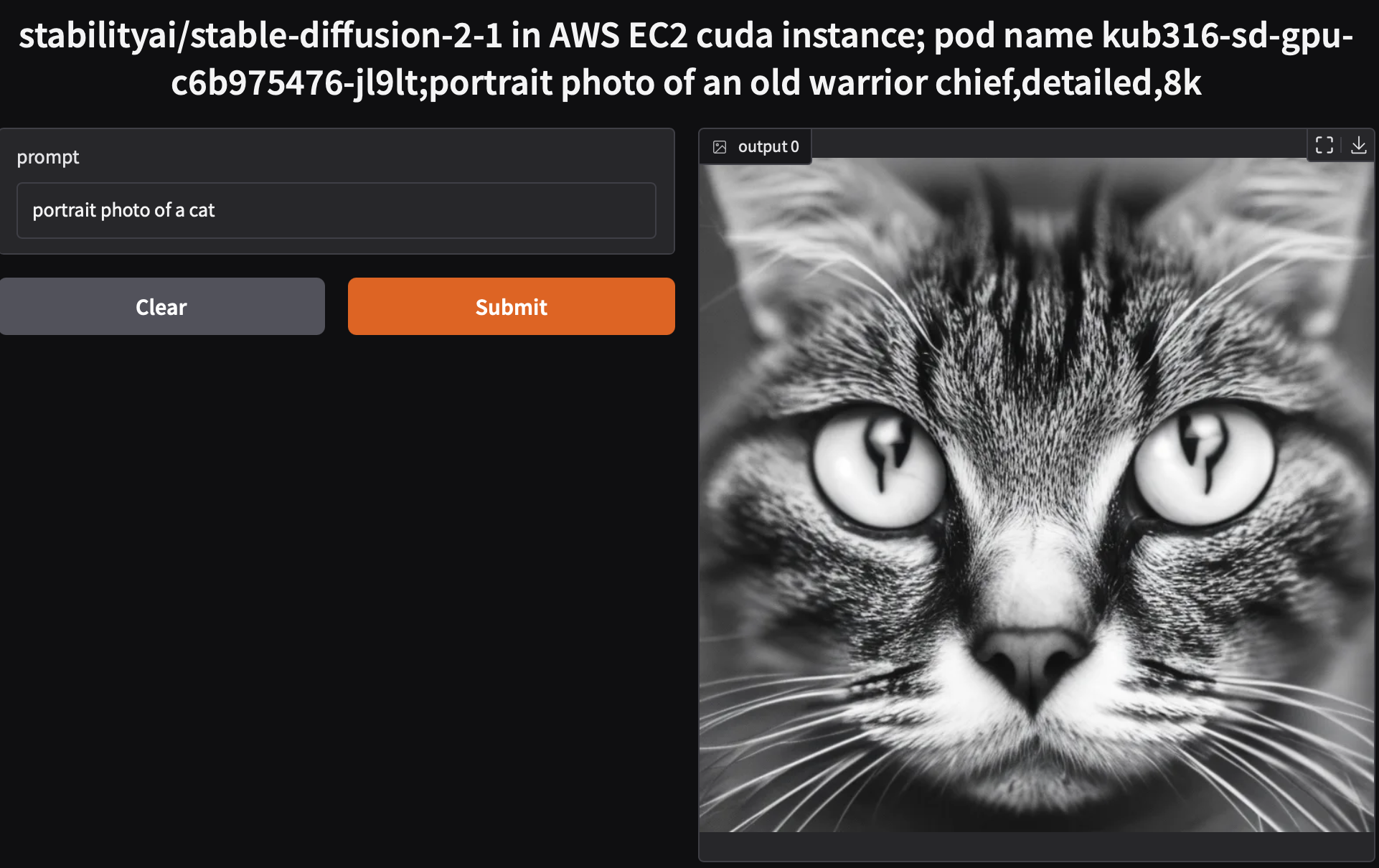}
        \caption{Portrait of a cat, GPU-based sample. Despite non-deterministic generation, quality is high and stylistically coherent.}
        \label{fig:gpu-gradio-sample}
    \end{subfigure}
    
    \vspace{0.5cm} 

    \begin{subfigure}[t]{\linewidth}
        \centering
        \includegraphics[width=\columnwidth]{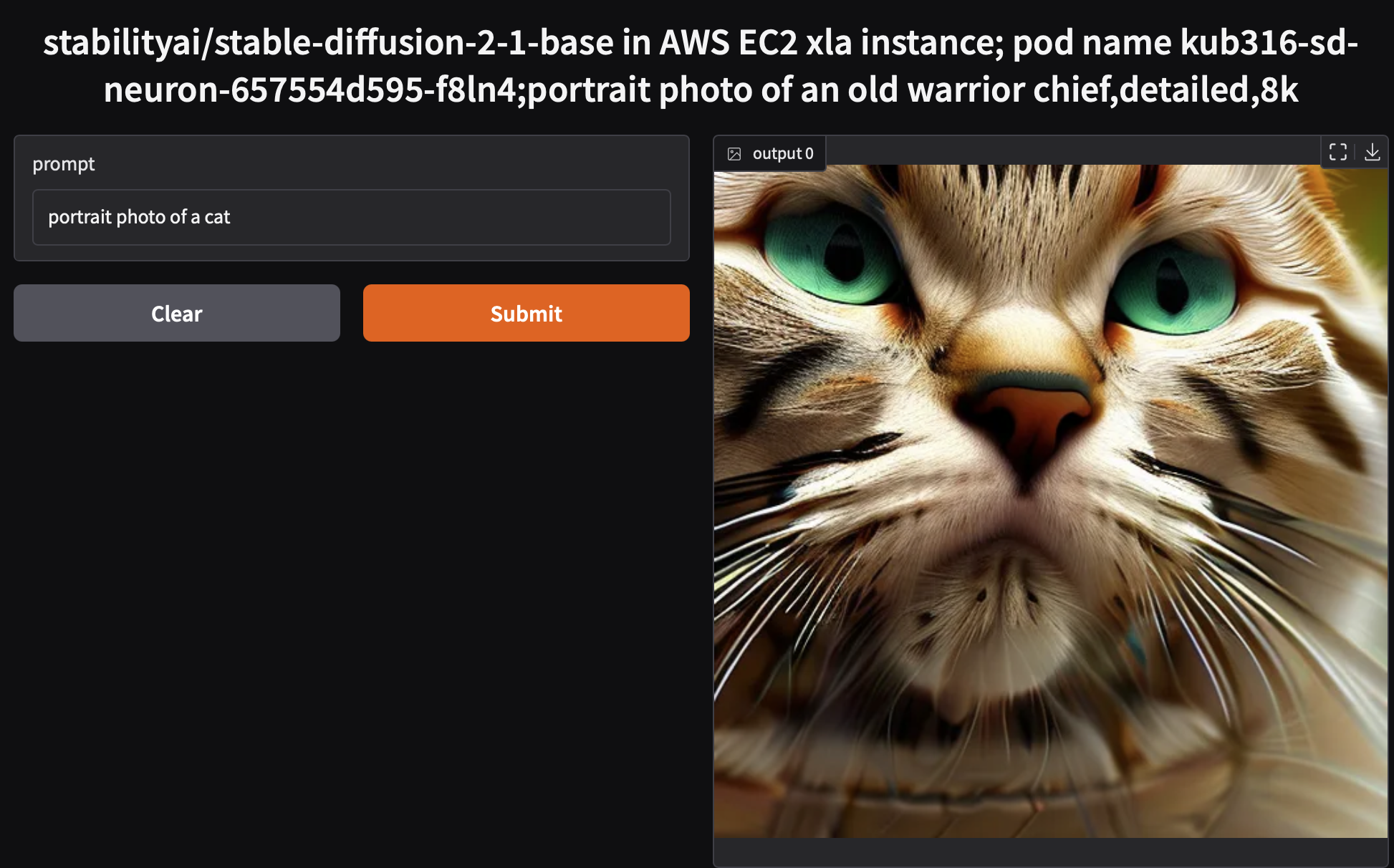}
        \caption{Portrait of a cat, Neuron-based sample. Although details differ slightly, quality and thematic fidelity remain consistent.}
        \label{fig:inf-gradio-sample}
    \end{subfigure}
    \caption{Model quality baseline under non-determinism}
    \label{fig:combined-gradio-sample}
\end{figure}

Non-deterministic models, particularly generative ones, can produce slightly outputs under identical inputs. To assess output quality consistency, we used a stylized image-generation model to produce two separate portraits of an ``old warrior.'' Figures~\ref{fig:gpu-gradio-sample} and \ref{fig:inf-gradio-sample} display the results.

While subtle variations in facial expression and background texture exist, the overarching style and fidelity remain comparable. Thus, the system’s design ensures that non-deterministic outputs maintain a consistent quality standard, irrespective of the underlying hardware resources or scaling decisions.

\subsection{Compute accelerator baseline comparisons}
\begin{figure}[ht]
    \centering
    \begin{subfigure}[t]{\linewidth}
        \centering
        \includegraphics[width=\columnwidth]{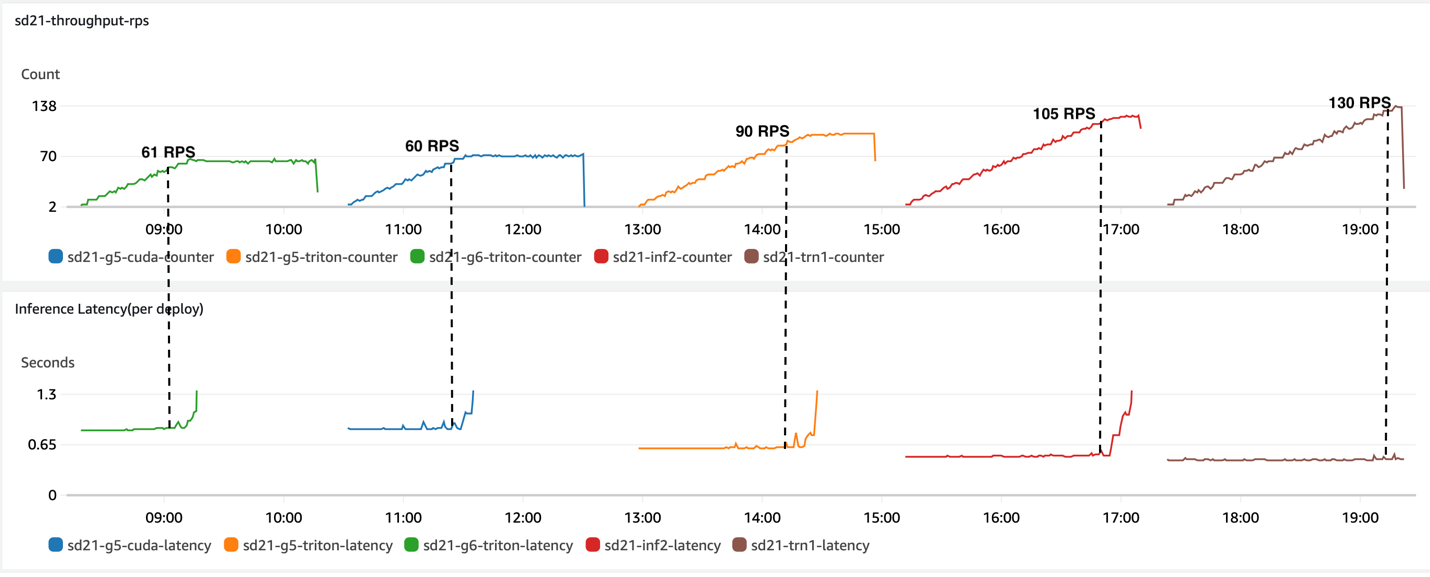}
        \caption{Inference latency, $L_i$, and throughput, $T_i$ per deployment unit $DU_i$(model-device-framework)}
        \label{fig:breakpoint-latency}
    \end{subfigure}
    
    \vspace{0.5cm} 

    \begin{subfigure}[t]{\linewidth}
        \centering
        \includegraphics[width=\columnwidth]{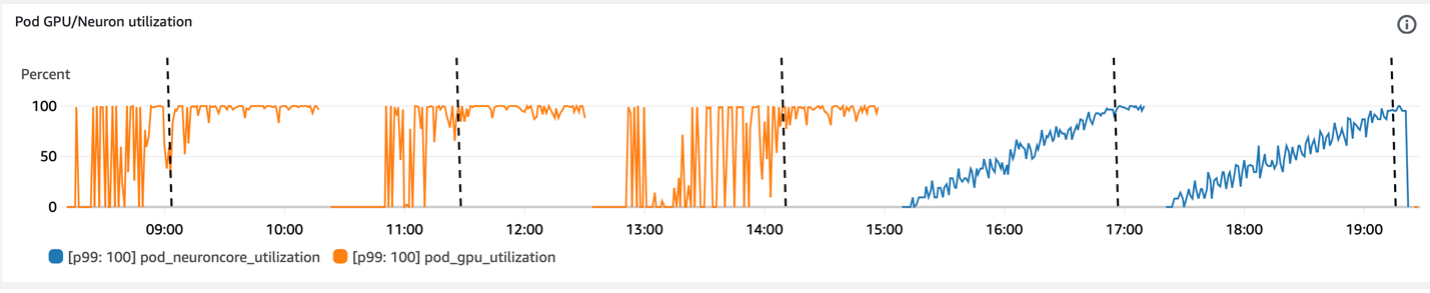}
        \caption{Compute accelerator utilization during load (neuron-core and GPU core)}
        \label{fig:figure5-breakpoint-util}
    \end{subfigure}
    \caption{look for the breaking point—when latency exceeds the set thresholds—on models loaded on Neuron and NVIDIA accelerators (Figure 4) or when the compute usage reaches over 80\%. We load test the application for each compute accelerator and framework combination, such as Inf2, Trn1, or GPU with CUDA, NeuronX, or Triton. The results define the $N^{modelProcessed}_i(t)$ that the autoscaler, KEDA, uses to scale the required number of $DU^p_i$ for each deployment combination. The breaking point occurs when throughput plateaus and latency exceeds 900 milliseconds. Below are the load tests conducted on A10G, L4 NVIDIA cores, and Inf2 and Trn1 Neuron cores.}
    \label{fig:combined-breakpoint}
\end{figure}

As a baseline, we first measured the performance of a single core loaded with a model with no autoscaling to determine the scale-out thresholds for each deployment unit $DU$ $DU^p_i$. This threshold is determined through experiments and observations, where we specifically look for the breaking point—when latency exceeds the set thresholds—on models loaded on Neuron and NVIDIA accelerators (Figure 4) or when the compute usage reaches over 80\% (Figure 5). We load test the application for each compute accelerator and framework combination, such as Inf2, Trn1, or GPU with CUDA, NeuronX, or Triton \cite{nvidia_triton}. The results define the targetMetricValue that KEDA uses to scale the required number of $DU^p_i$ for each deployment combination. The breaking point occurs when throughput plateaus and latency exceeds 900 milliseconds and accelrates beyind acceptable threshold. Below are the load tests conducted on A10G, L4 NVIDIA cores, and Inf2 and Trn1 Neuron cores with Stable-Diffusion 2.1\cite{stabilityaisd}. We skipped the CUDA eager mode with L4 NVIDIA, as it did not meet the minimum latency requirements. Figure~\ref{fig:breakpoint-latency} compares this baseline to our proposed approach, which dynamically leverages a mix of accelerators (GPUs and Inferentia/Trainium) as needed.

We calculate cost of inference per second for each deployment unit based on the breaking points and the Amazon EC2 on-demand pricing in Table \ref{tab:deployment_comparison}.
\begin{table}[htbp]
    \centering
    \resizebox{\columnwidth}{!}{%
    \begin{tabular}{|l|l|c|c|}
        \hline
        $DU$       & Cost of Compute/Hour & $T^{max}_i$ & Cost of Inference/Second \\ \hline
        (sd21, inf2, neuron)  & inf2.xlarge / \$0.7582      & 105        & 0.00733                    \\ \hline
        (sd21, trn1, neuron)  & trn1.2xlarge / \$1.3438     & 130        & 0.01023                    \\ \hline
        (sd21, g5, triton)    & g5.xlarge / \$1.0060        & 90         & 0.01118                    \\ \hline
        (sd21, g6, triton)    & g6.xlarge / \$0.8048        & 61         & 0.01320                    \\ \hline
        (sd21, g5, cuda)      & g5.xlarge / \$1.0060        & 60         & 0.01677                    \\ \hline
    \end{tabular}
    }
    \caption{Comparison of Deployment Units, Costs, and Throughput}
    \label{tab:deployment_comparison}
\end{table}

\subsection{Scaling Dynamics and Elasticity}
Our primary goal is to optimize the cost of inference at scale while maintaining sufficient compute capacity to meet user demands. To achieve this, we introduced two compute allocation and traffic distribution regimes, as described in Section \ref{subsec:capdynamics}. Our experiment distributes traffic across multiple deployment units $DU_i$ based on two key constraints: the Cost-Optimized Weight (Equation \ref{eq:throughputconstraint}) and the Capacity-Optimized Weight (Equation \ref{eq:capacityconstraint}).

In practice, user inference requests are efficiently handled by a cost-optimized pool, such as Inf2, while capacity-optimized pools serve as a fallback when demand exceeds the available resources in the cost-optimized pool. This approach ensures a balance between efficiency and reliability. Managing continuous weight distributions across multiple pools—especially when dealing with five or more—introduces unnecessary complexity with limited performance gains. Therefore, by categorizing pools into two types, the system enhances scalability, simplifies resource management, and maintains predictable behavior.

\subsubsection{Compute cost optimized configuration}
\begin{figure}[H]
    \centering
    
    \begin{subfigure}[t]{\linewidth}
        \centering
        \includegraphics[width=\columnwidth]{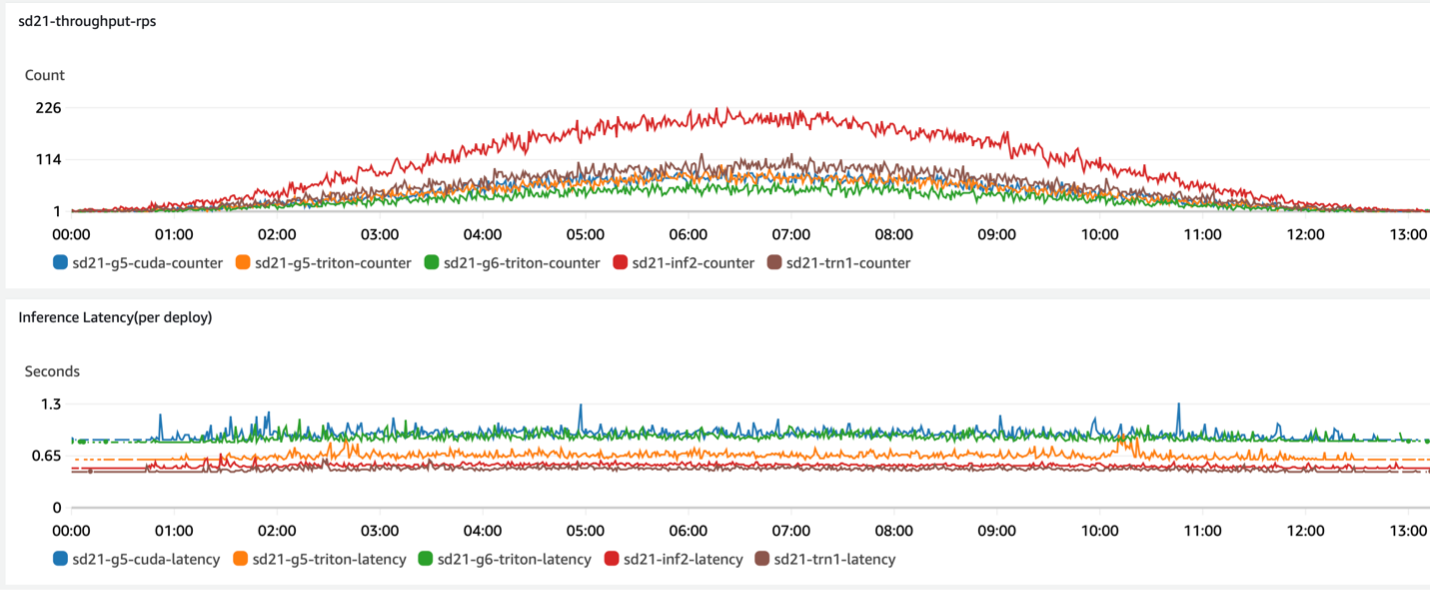}
        \caption{Cost optimized inference throughput, $T_i$ and latency, $L_i$, per deployment unit $DU_i$(model-device-framework)}
        \label{fig:cost-optimized-deploy}
    \end{subfigure}
    
    \vspace{0.5cm} 

    \begin{subfigure}[t]{\linewidth}
        \centering
        \includegraphics[width=\columnwidth]{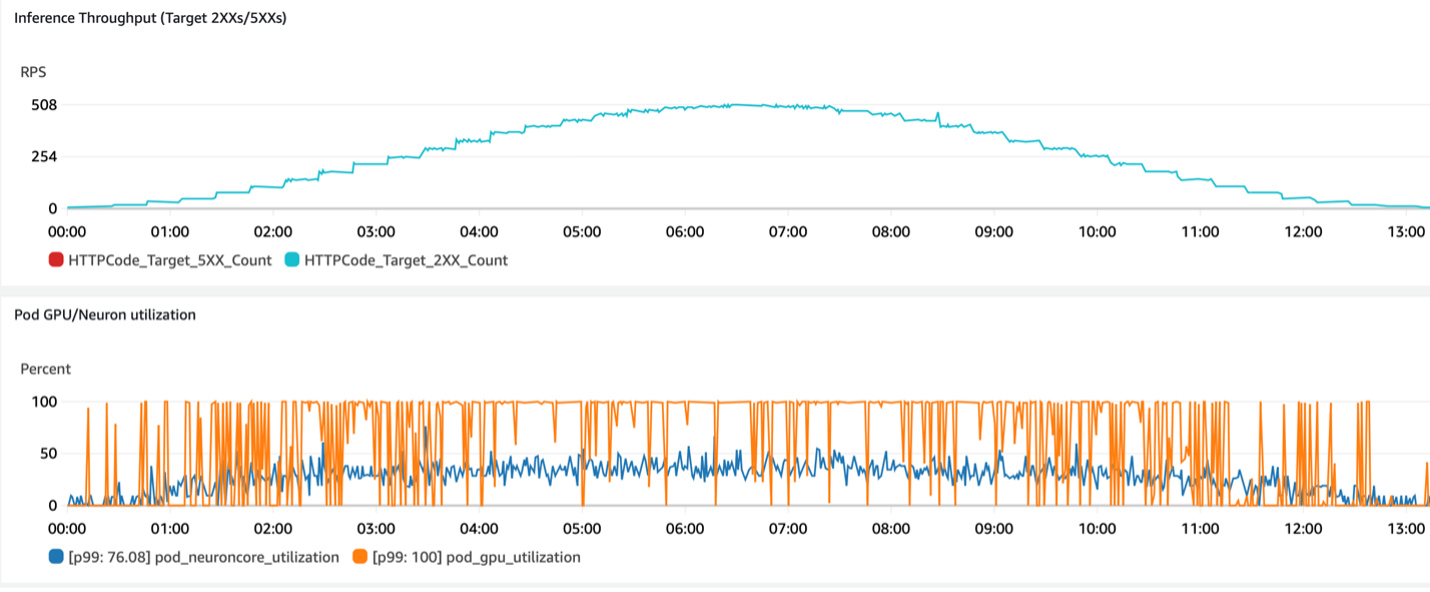}
        \caption{Cost optimized Compute accelerator utilization during load (neuron-core and GPU core)}
        \label{fig:cost-optimized-throughput}
    \end{subfigure}
    \captionsetup{font=small}
    \caption{The top graph shows the throughput (requests per second) for different deployment units over time, indicating a peak around mid-experiment, with the sd21-inf2-counter having the highest throughput. The bottom graph presents the inference latency per deployment, where the cost-optimized deployments maintain consistently low latency, while others show slightly higher variations. The top graph displays the total inference throughput, where successful requests (2XX) increase steadily and then plateau, with minimal error responses (5XX). The bottom graph depicts GPU/Neuron utilization, highlighting that while GPU utilization is consistent, Neuron utilization fluctuates significantly throughout the experiment.}
    \label{fig:combined-cost-optimized}
\end{figure}

The compute cost optimized configuration results illustrate the optimal compute allocation based on inference cost. The $sd21-inf2$ deployment handled 40\% of total requests with minimal latency, while the remaining deployments were allocated per the ALB ingress configuration. Figure \ref{fig:combined-cost-optimized} displays effective throughput, indicated by HTTP code 200 (successful requests) and HTTP code 500 (failures), while maintaining optimal utilization levels—70\% for Neuron cores and 90\% for GPU cores.
 
\FloatBarrier
\subsubsection{Compute capacity optimized configuration}
The weight for each capacity-optimized deployment unit is based on compute availability. If a deployment unit has available capacity, it receives a share of the traffic in a round-robin manner with the other available units. This weight ensures that traffic is distributed across all available capacity-optimized units, giving equal share to each available unit. When compute capacity is limited, your inference system must continue serving user requests while minimizing latency and maximizing application availability. To achieve this, we normalized the throughput of deployment units listed in Table \ref{tab:deployment_comparison} and consolidated the load balancer ingress into a single service that uses round-robin to distribute requests across deployment units, allocating resources evenly based on available capacity. The adjusted throughput per deployment unit aims to approximate uniformity, enabling nearly equal throughput across units as the load balancer distributes requests in a round-robin fashion. However, throughput also factors in both maximum and average latency per unit, allowing faster options like sd21-inf2 and sd21-trn1 to handle a higher volume of requests when possible. We used the deployment unit maximum throughput (Table \ref{tab:deployment_comparison}) and observed latency to calculate the target throughput as follow:

\begin{equation}
T^{\text{target}} = \frac{\sum_{i=1}^{n} T^{\text{max}}_i}{n}
\end{equation}

Then for each deployment unit, we set (Table \ref{tab:deployment_unit_performance}) the adjusted throughput, $T^{adjusted}$ to the minimum of either:

\begin{equation}
T^{\text{adjusted}}_i = \min\left(T_i, T^{\text{max}}_i\right), \quad \forall i \in \{1, 2, 3, 4, 5\}
\end{equation}

\begin{figure}[H]
    \centering
    \begin{subfigure}[t]{\linewidth}
        \centering
        \includegraphics[width=\columnwidth]{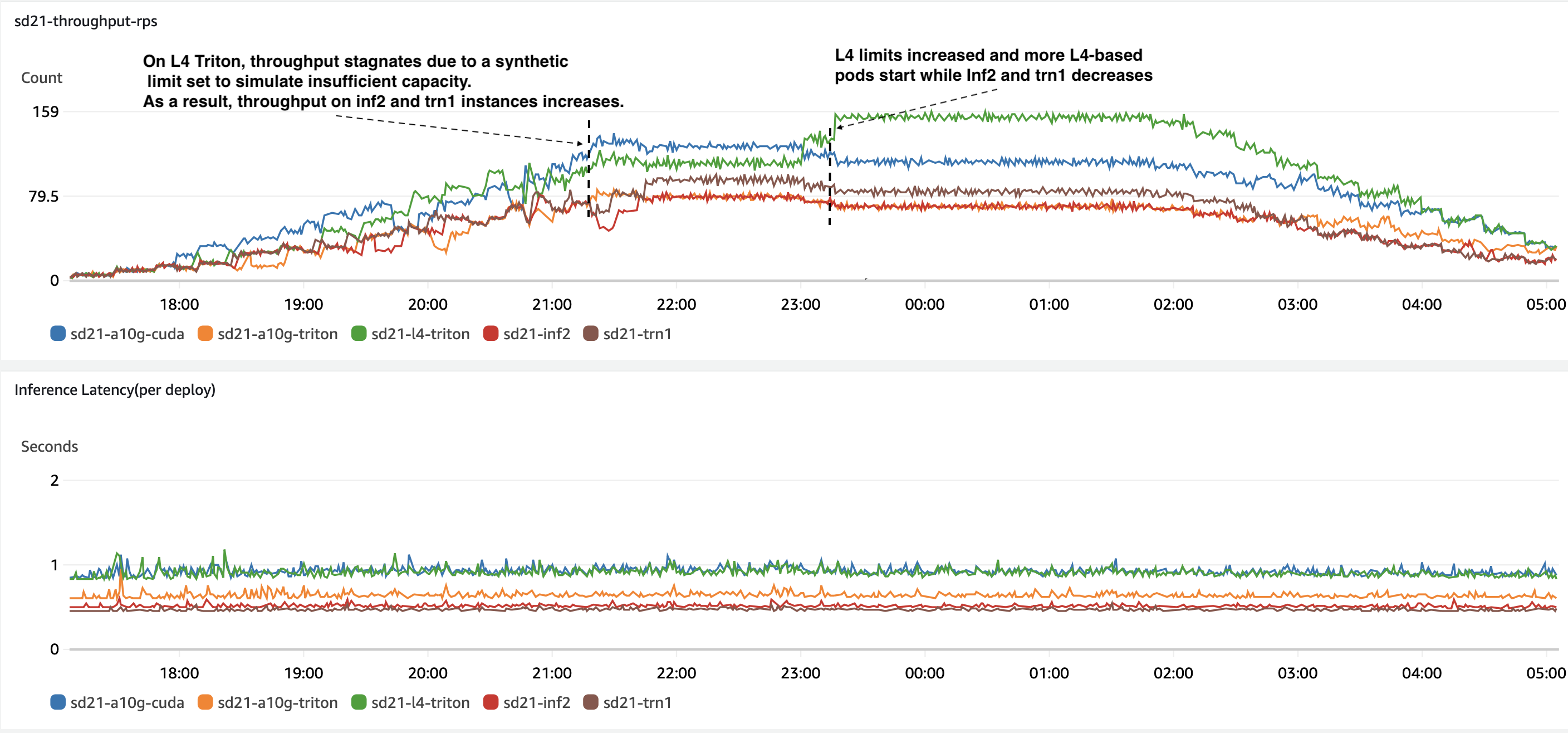}
        \caption{Capacity optimized deployment with equal round-robin load balancing}
        \label{fig:capacity-optimized-deploy}
    \end{subfigure}
    
    \vspace{0.5cm} 

    \begin{subfigure}[!ht]{\linewidth}
        \centering
        \includegraphics[width=\columnwidth]{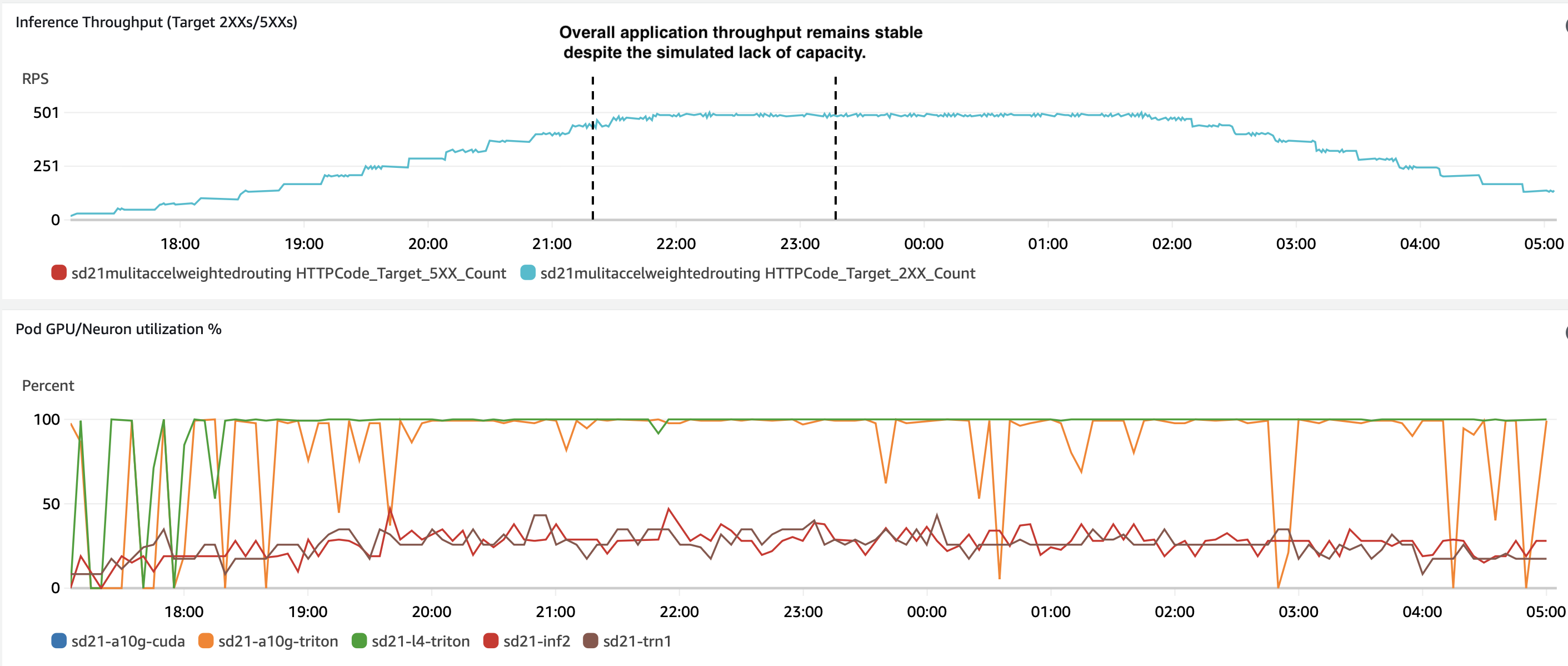}
        \caption{Capacity optimized deployment HTTP throughput and compute usage}
        \label{fig:capacity-optimized-throughput}
    \end{subfigure}
    \captionsetup{font=small}
    \caption{Initially, throughput on L4 instances stagnates due to a synthetic capacity limit, causing increased load on Inf2 and Trn1 instances to meet demand. After the L4 limits are raised, more L4-based pods are deployed, resulting in throughput stabilization and reduced load on Inf2 and Trn1. Despite the simulated constraint, overall application throughput remains stable, indicating the system’s robustness. Inference latency remains consistently low across all deployment units, while pod utilization shows stable usage for Inf2 and Trn1, with L4 utilization fluctuating due to scaling adjustments.}
    \label{fig:combined-capacity-optimized}
\end{figure}

\begin{table}[htbp]
    \centering
    \resizebox{\columnwidth}{!}{%
    \begin{tabular}{|l|c|c|c|c|}
        \hline
        $DU$ & $L$ (sec) & $T^{max}$ (RPS) & Cost of Inference/Second & $T^{adjusted}$ \\ \hline
        (sd21, inf2, neuron)  & 0.67 & 105 & 0.00733 & 89.2 \\ \hline
        (sd21, trn1, neuron)  & 0.51 & 130 & 0.01023 & 89.2 \\ \hline
        (sd21, g5, triton)    & 0.68 & 90  & 0.01118 & 89.2 \\ \hline
        (sd21, g6, triton)    & 0.96 & 61  & 0.01320 & 61.0 \\ \hline
        (sd21, g5, cuda)      & 0.92 & 60  & 0.01677 & 60.0 \\ \hline
    \end{tabular}
    }
    \caption{Performance metrics of different deployment units.}
    \label{tab:deployment_unit_performance}
\end{table}
\FloatBarrier

\subsubsection{Putting it all together - Failover to compute-optimized configuration with fallback to cost-optimized}
A binary step function is employed to prioritize cost-optimized pools, thereby minimizing operational costs. Meanwhile, the fallback mechanism ensures sufficient throughput during periods of high demand. Our controller continuously monitors the capacity pools (Karpenter NodePool) and switches between the binary steps based on the temporal capacity state relative to user demand.

\begin{figure}[ht]
    \centering
    \includegraphics[width=\linewidth]{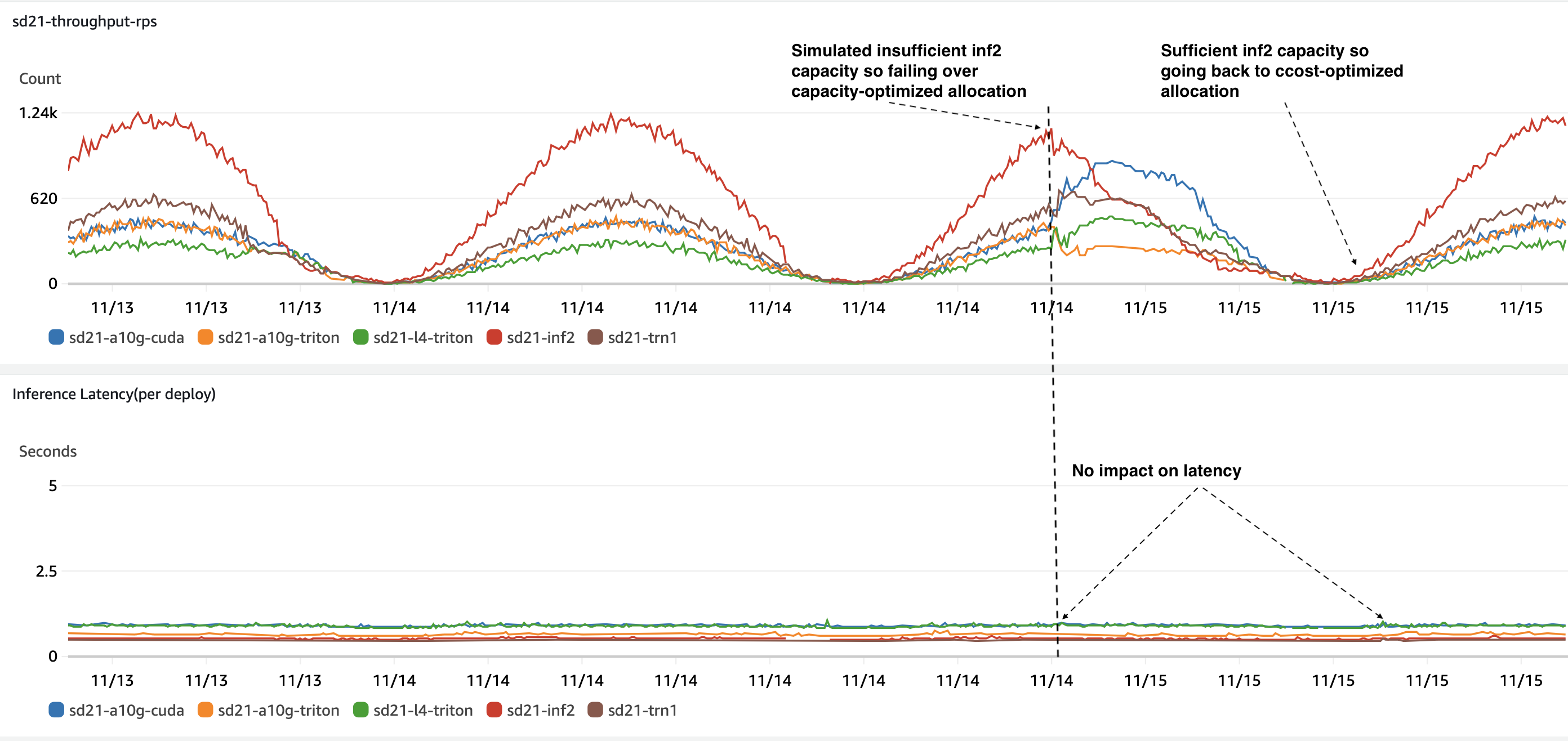}
    \captionsetup{font=small}
    \caption{The upper graph in the image shows "sd21-throughput-rps" (requests per second) across different compute configurations over time. The lower graph shows the latency. The RPS represents throughput for Stable Diffusion 2.1 (sd21) inference. The upper graph represents throughput, with measurements in increments up to approximately 1.24k requests per second. It highlights the system's ability to balance throughput by dynamically switching between cost- and capacity-optimized configurations, while keeping the latency low, ensuring high availability and efficient resource utilization.}
    \label{fig:opt23}
\end{figure}

Figure \ref{fig:opt23} summarizes the experiment in which a simulation of insufficient capacity for Inf2 was conducted on 11/14, prompting the controller to initiate a failover to the compute-optimized configuration to maintain throughput. The controller seamlessly transitioned to this configuration, managing the remaining load of the cycle without impacting latency, as evidenced by the consistent inference performance. At the beginning of the next wave on 11/15, when sufficient Inf2 capacity became available, the controller detected this change and automatically reverted to the cost-optimized allocation. This fallback ensured that the system could capitalize on cost savings while meeting performance requirements, as reflected in the stable throughput and controlled resource utilization.

\FloatBarrier
\section*{Conclusion}
In this work, we presented a scalable, hardware-agnostic framework for optimizing inference across heterogeneous compute accelerators, including NVIDIA GPUs, AWS Inferentia, and Trainium. By integrating Kubernetes-based orchestration tools and employing dynamic traffic distribution strategies, the framework effectively balances cost and capacity constraints. Our evaluation of Stable Diffusion demonstrated that the system maintains consistent throughput and low latency under varying load conditions while minimizing operational costs. The use of cost-optimized and capacity-optimized configurations, combined with an adaptive failover mechanism, ensured high availability and resource efficiency. This approach provides a practical pathway for deploying large-scale generative AI workloads in cloud environments, offering flexibility and reliability across diverse hardware setups. Future work will explore expanding the framework’s compatibility with additional accelerators and further refining its scaling and resource allocation strategies. For further exploration, the implementation of the framework and examples can be accessed at the GitHub repository: \href{https://github.com/aws-samples/scalable-hw-agnostic-inference}{https://github.com/aws-samples/scalable-hw-agnostic-inference}.

\printbibliography

@online{oci_image_spec,
  author       = {{Open Containers Initiative}},
  title        = {OCI Image Format Specification},
  url          = {https://github.com/opencontainers/image-spec},
  note         = {Version 1.0.2},
  year         = {2023},
  urldate      = {2025-01-11}
}

@inproceedings{pytorch,
  title     = {Automatic differentiation in {PyTorch}},
  author    = {Paszke, Adam and Gross, Sam and Massa, Francisco and Lerer, Adam and Bradbury, James and Chanan, Gregory and Killeen, Trevor and Lin, Zeming and Gimelshein, Natalia and Antiga, Luca and Desmaison, Alban and Kopf, Andreas and Yang, Edward and DeVito, Zachary and Raison, Martin and Tejani, Alykhan and Chilamkurthy, Sasank and Steiner, Benoit and Fang, Lu and Bai, Junjie and Chintala, Soumith},
  booktitle = {Proceedings of the 31st Conference on Neural Information Processing Systems (NIPS)},
  year      = {2017}
}

@online{torchdynamo,
  author  = {PyTorch Team},
  title   = {TorchDynamo: A Python-level JIT compiler designed to make unmodified PyTorch programs fast},
  url     = {https://pytorch.org/docs/stable/torch.compiler_dynamo_overview.html},
  year    = {2023},
  urldate = {2025-01-11}
}

@online{nvidia_triton,
  author  = {NVIDIA Corporation},
  title   = {NVIDIA Triton Inference Server},
  url     = {https://github.com/triton-inference-server/server},
  urldate = {2025-01-11},
  year    = {2023}
}

@online{stabilityaisd,
  author       = {Stability AI},
  title        = {Stable Diffusion Model},
  url          = {https://github.com/Stability-AI/stablediffusion},
  note         = {Accessed: 2025-01-11},
  year         = {2022}
}

@online{kubernetes,
  author  = {The Kubernetes Authors},
  title   = {Kubernetes Documentation},
  url     = {https://kubernetes.io/docs/home/},
  urldate = {2025-01-11},
  year    = {2023}
}

@online{autoscaler_tool,
  author  = {KEDA Authors},
  title   = {Kubernetes Event-driven Autoscaling (KEDA)},
  url     = {https://keda.sh/},
  urldate = {2025-01-11},
  year    = {2023}
}

@online{node_provisioner_tool,
  author  = {Karpenter Authors},
  title   = {Karpenter Documentation},
  url     = {https://karpenter.sh/},
  urldate = {2025-01-11},
  year    = {2023}
}

@online{huggingface_repos,
  author  = {Hugging Face},
  title   = {Hugging Face Model Hub},
  url     = {https://huggingface.co/models},
  urldate = {2025-01-11},
  year    = {2023}
}

@online{awsneuron,
  author  = {Amazon Web Services},
  title   = {AWS Neuron SDK Documentation},
  url     = {https://awsdocs-neuron.readthedocs-hosted.com/},
  urldate = {2025-01-11},
  year    = {2023}
}

@online{aws_trainium_systolic,
  author       = {Amazon Web Services},
  title        = {Amazon EC2 Trn1 Instances - Trainium Systolic Array Architecture Explained},
  url          = {https://aws.amazon.com/ec2/instance-types/trn1/},
  urldate      = {2025-01-12},
  year         = {2024},
  note         = {Accessed: 2025-01-12}
}

@online{nvidia_hopper_arch,
  author       = {NVIDIA Corporation},
  title        = {NVIDIA Hopper Architecture In-Depth},
  url          = {https://www.nvidia.com/en-us/data-center/hopper-gpu-architecture/},
  urldate      = {2025-01-12},
  year         = {2024},
  note         = {Accessed: 2025-01-12}
}

@article{k8shpc,
  author    = {He, Kai and Ellsworth, Michael and others},
  title     = {Kubernetes-Based Orchestration for HPC: A Comparative Study on Multi-Accelerator Clusters},
  journal   = {Proceedings of SC '24},
  pages     = {1--12},
  year      = {2024},
  note      = {DOI: 10.1109/SC.2024.00001}
}

@article{cost_efficient_gpu_cluster,
  author = {Yu, Ruofan and Wang, Qian and Jia, Ning and Mao, Yuyi and Liang, Yun and Xu, Chengzhong},
  title = {Cost Efficient GPU Cluster Management for Training and Inference Workloads},
  journal = {Energies},
  volume = {15},
  number = {2},
  pages = {474},
  year = {2022},
  publisher = {MDPI},
  doi = {10.3390/en15020474},
  url = {https://www.mdpi.com/1996-1073/15/2/474},
  urldate = {2025-03-24}
}

@inproceedings{heterogeneous_edge_eval,
  author = {Cervantes, Abelardo and Varghese, Blesson and Boubekeur, Tamy},
  title = {Evaluating the Cost-Effectiveness of Heterogeneous Edge Platforms},
  booktitle = {Proceedings of the 15th ACM Workshop on Hot Topics in Edge Computing (HotEdge)},
  pages = {1--7},
  year = {2023},
  publisher = {ACM},
  doi = {10.1145/3583740.3628437},
  url = {https://dl.acm.org/doi/10.1145/3583740.3628437},
  urldate = {2025-03-24}
}

@online{aws_ml_optimization,
  author = {Amazon Web Services},
  title = {Optimizing Machine Learning Inference on AWS},
  year = {2022},
  url = {https://pages.awscloud.com/rs/112-TZM-766/images/25%20March%20-%20ML%20Fridays%20-%20Optimizing%20machine%20learning%20inference%20on%20AWS.pdf},
  urldate = {2025-03-24},
  note = {Accessed: 2025-03-24}
}

@inproceedings{vaswani2017attention,
  title     = {Attention Is All You Need},
  author    = {Vaswani, Ashish and Shazeer, Noam and Parmar, Niki and Uszkoreit, Jakob and Jones, Llion and Gomez, Aidan N. and Kaiser, {\L}ukasz and Polosukhin, Illia},
  booktitle = {Proceedings of the 31st International Conference on Neural Information Processing Systems (NeurIPS)},
  year      = {2017},
  url       = {https://arxiv.org/abs/1706.03762}
}

@article{brown2020language,
  title     = {Language Models are Few-Shot Learners},
  author    = {Brown, Tom B. and Mann, Benjamin and Ryder, Nick and Subbiah, Melanie and Kaplan, Jared and Dhariwal, Prafulla and Neelakantan, Arvind and Shyam, Pranav and Sastry, Girish and Askell, Amanda and others},
  journal   = {arXiv preprint arXiv:2005.14165},
  year      = {2020},
  url       = {https://arxiv.org/abs/2005.14165}
}

@article{dosovitskiy2021vit,
  title     = {An Image is Worth 16x16 Words: Transformers for Image Recognition at Scale},
  author    = {Dosovitskiy, Alexey and Beyer, Lucas and Kolesnikov, Alexander and Weissenborn, Dirk and Zhai, Xiaohua and Unterthiner, Thomas and Dehghani, Mostafa and Minderer, Matthias and Heigold, Georg and Gelly, Sylvain and Uszkoreit, Jakob and Houlsby, Neil},
  journal   = {International Conference on Learning Representations (ICLR)},
  year      = {2021},
  url       = {https://arxiv.org/abs/2010.11929}
}

@inproceedings{wolf2020transformers,
  title     = {Transformers: State-of-the-Art Natural Language Processing},
  author    = {Wolf, Thomas and Debut, Lysandre and Sanh, Victor and Chaumond, Julien and Delangue, Cl{\'e}ment and Moi, Anthony and Cistac, Pierric and Rault, Tim and Louf, R{\'e}mi and Funtowicz, Morgan and Brew, Joe},
  booktitle = {Proceedings of the 2020 Conference on Empirical Methods in Natural Language Processing: System Demonstrations},
  pages     = {38--45},
  year      = {2020},
  doi       = {10.18653/v1/2020.emnlp-demos.6},
  url       = {https://aclanthology.org/2020.emnlp-demos.6}
}

@inproceedings{paszke2019pytorch,
  title     = {PyTorch: An Imperative Style, High-Performance Deep Learning Library},
  author    = {Paszke, Adam and Gross, Sam and Massa, Francisco and Lerer, Adam and Bradbury, James and Chanan, Gregory and Killeen, Trevor and Lin, Zeming and Gimelshein, Natalia and Antiga, Luca and others},
  booktitle = {Proceedings of the 33rd Conference on Neural Information Processing Systems (NeurIPS)},
  year      = {2019},
  url       = {https://papers.nips.cc/paper_files/paper/2019/hash/bdbca288fee7f92f2bfa9f7012727740-Abstract.html}
}

@online{aws_neuron_pytorch,
  author    = {{AWS Neuron SDK Team}},
  title     = {PyTorch Neuron — Neuron SDK Documentation},
  url       = {https://awsdocs-neuron.readthedocs-hosted.com/en/latest/frameworks/torch/torch-neuronx-guide.html},
  note      = {Accessed 2025-03-25},
  year      = {2024}
}

@inproceedings{moritz2018ray,
  title     = {Ray: A Distributed Framework for Emerging AI Applications},
  author    = {Moritz, Philipp and Nishihara, Robert and Wang, Stephanie and Tumanov, Alexey and Liaw, Richard and Liang, Eric and Elibol, Melih and Yang, Zongheng and Paul, William and Jordan, Michael I. and Stoica, Ion},
  booktitle = {13th USENIX Symposium on Operating Systems Design and Implementation (OSDI)},
  year      = {2018},
  pages     = {561--577},
  url       = {https://www.usenix.org/system/files/osdi18-moritz.pdf}
}

@inproceedings{kwon2023vllm,
  title     = {Efficient Memory Management for Large Language Model Serving with PagedAttention},
  author    = {Kwon, Woosuk and Li, Zhuohan and Zhuang, Siyuan and Sheng, Ying and Zheng, Lianmin and Zhang, Hao and Shao, Yuxin and Gonzalez, Joseph E. and Zaharia, Matei and Stoica, Ion},
  booktitle = {Proceedings of the 29th ACM Symposium on Operating Systems Principles (SOSP)},
  year      = {2023},
  url       = {https://arxiv.org/abs/2309.06180}
}

@online{vllm_neuron_support,
  author    = {{vLLM Team}},
  title     = {vLLM Neuron Support — vLLM Documentation},
  url       = {https://docs.vllm.ai/en/latest/usage/accelerators.html#aws-neuron},
  note      = {Version 0.3.3 — Neuron support for Inferentia and Trainium},
  year      = {2024},
  urldate   = {2025-03-25}
}

\end{document}